\documentstyle[amsmath,epsf,a4,11pt]{article}

\date{}

\begin{document}

\title{Geometrical Aspects of Cosmic Magnetic Fields}

\author{Christos G. Tsagas\footnote{Present address: Department of
Mathematics and Applied Mathematics, University of Cape Town,
Rondebosch 7701, South Africa.}\\{\small Relativity and Cosmology
Group, University of Portsmouth, Portsmouth PO1~2EG, England}}

\maketitle

\begin{abstract}
We discuss how the vector nature of magnetic fields, and the
geometrical interpretation of gravity introduced by general
relativity, lead to a special coupling between magnetism and
spacetime curvature. This magneto-geometrical interaction
effectively transfers the tension properties of the field into the
spacetime fabric, triggering a variety of effects with potentially
far-reaching implications.\\

\end{abstract}

\section{Introduction}
It has long been thought that magnetic fields might have played a
role during the formation and the evolution of the observed large
scale structure \cite{RR}. Recently, this idea has received
renewed interest manifested by the increasing number of related
papers that have appeared in the literature \cite{KOR,TB}.
Nevertheless, there are still only a few fully relativistic
approaches available. Most treatments are either Newtonian or
semi-relativistic. As such, they are bound to exclude certain
features of the magnetic nature. Two key features are the vector
nature of the field and the tension properties of magnetic force
lines. In general relativity vectors have quite a different status
than ordinary scalar sources, such as the energy density and
pressure of matter. The geometrical nature of Einstein's theory
means that vector fields are directly coupled to the spacetime
curvature, as manifested by the Ricci identity
\begin{equation}
2\nabla_{[a}\nabla_{b]}B_c=R_{abcd}B^d\,,  \label{Ricci}
\end{equation}
applied here to the magnetic vector $B_a$, where $R_{abcd}$ is the
spacetime Riemann tensor. The Ricci identity plays a fundamental
role in the mathematical formulation of general relativity.
Essentially, it is the definition of spacetime curvature itself.
We call the special interaction reflected in the right hand side
of Eq.~(\ref{Ricci}) the {\em magneto-curvature coupling}. This
coupling goes beyond the standard interplay between matter and
geometry as introduced by the Einstein field equations. In fact,
it makes the magnetic field an inseparable part of the spacetime
fabric by effectively transferring its properties to the spacetime
itself. The key property appears to be the tension of the magnetic
lines of force.

Magnetic fields transmit stresses between regions of material
particles and fluids. The field exerts an isotropic pressure in
all directions and carries a tension along the magnetic lines of
force. Each small flux-tube behaves like an infinitely elastic
rubber band, while neighbouring tubes expand against each other
under their own pressure. Equilibrium exists only when a balance
between pressure and tension is possible. To unravel these tension
properties, consider the energy momentum tensor of a pure magnetic
field
\begin{equation}
T_{ab}={\textstyle{1\over2}}B^2u_au_b+
{\textstyle{1\over6}}B^2h_{ab}+ \Pi_{ab}\,,  \label{mTab}
\end{equation}
where $B^2=B_aB^a$ and $\Pi_{ab}=(B^2/3)h_{ab}-B_aB_b$. Thus, the
field behaves as an imperfect fluid with energy density $\rho_{\rm
m}=B^2/3$, isotropic pressure $p_{\rm m}=B^2/6$ and anisotropic
pressure $\Pi_{ab}$. The tension properties of the field are
incorporated in the eigenvalues of the symmetric trace-free tensor
$\Pi_{ab}$. Orthogonal to $B_a$ one finds two positive eigenvalues
equal $1/3$ each. Thus, the magnetic pressure perpendicular to the
field lines is positive, reflecting their tendency to push each
other apart. In the $B_a$ direction, however, the associated
eigenvalue is $-2/3$ and the magnetic pressure is negative. The
minus sign reflects the tension properties of the field lines,
namely their tendency to remain as `straight' as possible.

These elastic magnetic properties are effectively injected into
the spacetime fabric itself, through the aforementioned coupling
between magnetism and geometry. The result is a magneto-curvature
stress which always reacts to curvature distortions and tries to
restore spatial flatness~\cite{TM, MT}. This relativistic
magneto-curvature stress closely resembles the classical one
exerted by distorted magnetic field lines (see e.g.~\cite{P}). The
difference is that, in the relativistic case, the distortion of
the field pattern is triggered by the spacetime geometry itself.
The implications are widespread and far from trivial.

\section{The magnetic impact on universal expansion}
Consider a general spacetime filled with a magnetised, highly
conductive, perfect fluid. Its volume expansion is governed by the
non-linear Raychaudhuri equation \cite{TB}
\begin{equation}
{\textstyle{1\over3}}\Theta^2q=
{\textstyle{1\over2}}\left(\rho+3p+B^2\right)+
2\left(\sigma^2-\omega^2\right) -\nabla^aA_a- \Lambda\,,
\label{Ray}
\end{equation}
where $q$ is the deceleration parameter, $\sigma^2$ and $\omega^2$
are the shear and vorticity magnitudes respectively and $\Lambda$
is the cosmological constant. The state of the expansion is
determined by the sign of the right-hand side of Eq.~(\ref{Ray}).
Positive terms decelerate the universe while negative ones lead to
acceleration. Clearly, conventional matter and shear effects slow
the expansion down. On the other hand, vorticity and a positive
cosmological constant accelerate the universe. Hence, every term
on the right hand side of Eq.~(\ref{Ray}) has a clear kinematical
role with the exception of $\nabla^aA_a$. The latter can be either
positive or negative, depending on the specific form of the
4-acceleration. In our case $A_a$ obeys the non-linear Euler
formula given by~\cite{TB}
\begin{equation}
\left(\rho+p+{\textstyle{2\over3}}B^2\right)A_a=-c_{\rm s}^2{\rm
D}_a\rho- \varepsilon_{abc}B^b{\rm curl}B^c -A^b\Pi_{ba}\,,
\label{Euler}
\end{equation}
where $A_a$ is the fluid 4-acceleration, $c_{\rm
s}^2=\dot{p}/\dot{\rho}$ is the sound speed squared and
$\varepsilon_{abc}$ the spatial alternating tensor. In a weakly
magnetised, slightly inhomogeneous and anisotropic, almost-FRW
universe Eqs.~(\ref{Euler}) and (\ref{Ray}) linearise to give
\cite{MT}
\begin{equation}
{\textstyle{1\over3}}\Theta^2{\rm q}=
{\textstyle{1\over2}}\rho(1+3w)+ \frac{c_{\rm
s}^2\Delta}{(1+w)a^2}+ \frac{c_{\rm a}^2{\cal B}}{2(1+w)a^2}
-\frac{2kc_{\rm a}^2}{(1+w)a^2}- \Lambda\,. \label{Re}
\end{equation}
In the above $\Delta$ and ${\cal B}$ describe scalar perturbations
in the fluid and the magnetic energy density respectively, $c_{\rm
a}^2=B^2/\rho$ is the square of the Alfv\'en speed, $k=0\,,\pm1$
is the background curvature index and $a$ is the scale factor.
Given that in the linear regime the mean values of $\Delta$ and
${\cal B}$ are zero, one expects that on average Eq.~(\ref{Re})
looks like
\begin{equation}
{\textstyle{1\over3}}\Theta^2{\rm q}=
{\textstyle{1\over2}}\rho(1+3w)- \frac{2kc_{\rm a}^2}{(1+w)a^2}\,,
\label{Re1}
\end{equation}
where $\Lambda=0$ from now on. Note the magneto-curvature term in
the right-hand side which results from the coupling between
magnetism and geometry as manifested in Eq.~(\ref{Ricci}). This
term affects the expansion in two completely different ways
depending on the sign of the background curvature. In particular,
the magneto-geometrical effects slow the expansion down when
$k=-1$ but tend to accelerate the expansion if $k=+1$. Such a
behaviour seems odd, especially since positive curvature is always
associated with gravitational collapse. The explanation lies in
the elastic properties of the field lines. As curvature distorts
the magnetic force lines their tension backreacts giving rise to a
restoring magneto-curvature stress \cite{TM}. The magnetic
backreaction has kinematical, dynamical as well as geometrical
implications. In Eq.~(\ref{Re1}), for example, the tension of the
field adjusts the expansion rate of the universe to minimize the
kinematical effects of curvature. As a result the expansion rate
is brought closer to that of a flat FRW model. Overall, it looks
as though the elastic properties of the field have been
transferred into space. According to Eq.~(\ref{Re1}), the
magneto-curvature effects also depend on the material component of
the universe. When dealing with conventional matter (i.e.~for
$0\leq w\leq1$) the most intriguing cases occur in positively
curved spaces. In particular, when $w=1$ (i.e.~for stiff matter)
the Alfv\'en speed grows as $c_{\rm a}^2\propto a^2$ and the
magneto-curvature term in Eq.~(\ref{Re1}) becomes
time-independent. In this case the field acts as an effective
positive cosmological constant. For radiation and dust, on the
other hand, $c_{\rm a}^2={\rm const.}$ and $c_{\rm a}^2\propto
a^{-1}$ respectively. In these cases the magneto-curvature term is
no longer time independent but drops with time mimicking a
time-decaying quintessence~\cite{MT}.

The magneto-curvature effects discussed so far, subtle though they
may be, remain secondary unless the field is relatively strong.
However, the coupling between magnetism and geometry means that
even weak fields can have a strong overall impact when the
curvature is strong. To demonstrate how this might happen, we
consider a weakly magnetised spatially open cosmology.  For $k=-1$
Eq.~(\ref{Re1}) becomes
\begin{equation}
{\textstyle{1\over3}}\Theta^2{\rm q}=
{\textstyle{1\over2}}\rho(1+3w)+ \frac{2c_{\rm a}^2}{(1+w)a^2}\,,
\label{Re2}
\end{equation}
where now the magneto-curvature term tends to decelerate the
expansion. Let us assume a spacetime filled with non-conventional
matter, namely that $-1\leq w<0$. Scalar fields, for example, can
have an effective equation of state that satisfies this
requirement. Such models allow for an early curvature dominated
regime with $\Omega\ll1$. Given that $\rho\propto a^{-3(1+w)}$ and
$c_{\rm a}^2\propto a^{-1+3w}$, the magneto-curvature term in
Eq.~(\ref{Re1}) can dominate the early expansion, even when the
field is weak, if $-1\leq w\leq-1/3$. In this case the accelerated
inflationary phase, which otherwise would have been inevitable, is
suppressed. Instead of inflating the magnetised universe remains
in a state of decelerated expansion. For $w=-1$, in particular,
the mere presence of the field can inhibit the de Sitter
inflationary regime if $\Omega<0.5$ \cite{MT}. Clearly, this
result challenges the widespread perception that magnetic fields
are relatively unimportant for cosmology. Even weak fields can
play a decisive role when the curvature is strong. Moreover, it
casts doubt on the efficiency of standard inflation in the
presence of primordial magnetism.

\section{Magnetic fields and gravity waves}
Let us now turn our attention to geometry and examine the
implications of the magnetic presence for propagating
gravitational radiation. The production of gravity waves by
stochastic magnetic fields has been investigated in~\cite{CD}.
Here we will take a more geometrical approach and look at the
implications of the tension properties of the field for gravity
waves passing through a magnetised region. Our starting point is a
spatially flat FRW background universe filled with a highly
conductive perfect fluid. We will then perturb this background by
allowing for weak gravitational waves and a weak magnetic field.
Covariantly, gravity waves are described via the electric
($E_{ab}$) and the magnetic ($H_{ab}$) parts of the Weyl tensor
\cite{E}. Their magnitudes, $E^2=E_{ab}E^{ab}/2$ and
$H^2=H_{ab}H^{ab}/2$, provide a measure of the wave's energy
density and amplitude. Given that $H_{ab}={\rm curl}\sigma_{ab}$,
we can simplify the problem by replacing the magnetic Weyl tensor
with the shear. Note that the field couples to gravitational
radiation directly via the anisotropic magnetic stresses, which
affect the propagation of both $E_{ab}$ and $\sigma_{ab}$
\cite{MTU}. Having set the constraints that isolate tensor
perturbations in a magnetised universe (see \cite{MTU}), we arrive
at the system
\begin{eqnarray}
(E^2)^{.}&=&-2\Theta E^2- {\textstyle{1\over2}}\rho(1+w){\cal X}-
{\textstyle{1\over2}}\Theta B^2{\cal E}\,,\nonumber\\
(\sigma^2)^{.}&=&-{\textstyle{4\over3}}\Theta\sigma^2- {\cal X}
-{\textstyle{1\over2}}B^2\Sigma\,,\nonumber\\ \dot{{\cal X}}&=&-
{\textstyle{5\over3}}\Theta{\cal X}- 2E^2- \rho(1+w)\sigma^2-
{\textstyle{1\over2}}B^2{\cal E}-{\textstyle{1\over2}}\Theta
B^2\Sigma\,,  \label{dotSigma}\\ \dot{{\cal E}}&=&-\Theta{\cal E}-
{\textstyle{1\over2}}\rho(1+w)\Sigma- {\textstyle{1\over3}}\Theta
B^2\,,\nonumber\\
\dot{\Sigma}&=&-{\textstyle{2\over3}}\Theta\Sigma-{\cal
E}-{\textstyle{1\over3}}B^2\,,\nonumber
\end{eqnarray}
with $B^2\propto a^{-4}$, ${\cal X}=E_{ab}\sigma^{ab}$, ${\cal
E}=E_{ab}\eta^a\eta^b$ and $\Sigma=\sigma_{ab}\eta^a\eta^b$
($\eta_a=B_a/\sqrt{B^2}$). The last two scalars are related via
the Gauss-Codacci equation by
\begin{equation}
{\cal E}={\textstyle{1\over3}}\Theta\Sigma+
{\textstyle{1\over3}}B^2+ {\rm R}\,,  \label{GC}
\end{equation}
where ${\rm R}=[{\cal R}_{(ab)}-({\cal R}/3)h_{ab}]\eta^a\eta^b$
describes spatial curvature distortions in the direction of the
magnetic field lines. For radiation, the late-time solutions for
$E^2$ and $\sigma^2$ are \cite{MTU}
\begin{eqnarray}
E^2&=&{\textstyle{4\over9}}
\left[E_0^2+\frac{\sigma_0^2}{4t_0^2}-\frac{{\cal
X}_0}{2t_0}\right] \left(\frac{t_0}{t}\right)^2
-{\textstyle{2\over9}}\left({\textstyle{1\over6}}B_0^2+{\rm R}_0
\right)B_0^2 \left(\frac{t_0}{t}\right)^2\,, \nonumber\\
\label{LTS}\\ \sigma^2&=&{\textstyle{1\over9}}
\left[\sigma_0^2+4E_0^2t_0^2-2{\cal X}_0t_0\right]
-{\textstyle{2\over9}}\left({\textstyle{1\over6}}B_0^2+{\rm R}_0
\right)B_0^2t_0^2\,, \nonumber
\end{eqnarray}
with an analogous result for dust \cite{MTU}. Note that the terms
in square brackets determine the magnetic-free case. According to
(\ref{LTS}), the field leaves the evolution rates of both $E^2$
and $\sigma^2$ unchanged but modifies their magnitudes. The
magnetic impact depends on the initial conditions and it is
twofold. There is what one might call a pure magnetic effect,
independent of the spatial curvature, which suppresses the energy
of the wave. It becomes apparent when we set ${\rm R}_0=0$ in
Eqs.~(\ref{LTS}). Hence, when one starts from a FRW background, as
we did here, the field presence will have a smoothing effect on
the waves. Given the inherent weakness of gravitational radiation,
these magnetic effects are potentially detectable even when
relatively weak fields are involved. Solution (\ref{LTS}) also
reveals a magneto-curvature effect on gravitational radiation.
This is encoded in the ${\rm R}_0$-term and depends entirely on
the spatial curvature. For ${\rm R}_0>0$, namely when the
curvature distortion along the field lines is positive, the effect
is to decrease the energy of the wave. On the other hand, the
magneto-curvature term will increase the wave's energy when ${\rm
R}_0<0$. These magneto-geometrical effects result from the tension
properties of the field and get stronger with increasing curvature
distortion. Let us take a closer look at them. According to
Eq.~(\ref{GC}), the scalar ${\rm R}$ describes distortions in the
local spatial curvature generated by the propagating magnetised
gravity wave. The magneto-geometrical terms in (\ref{LTS}) modify
the energy density of the wave as though they are trying to
minimize the curvature distortion. In other words, the pure
magneto-curvature effect shows a tendency to preserve the spatial
flatness of the background universe. Earlier, an analogous
magneto-curvature effect was also observed on the expansion rate
of spatially curved FRW universes. This pattern of behaviour
raises the question as to whether it reflects a generic feature of
the magnetic nature. More specifically, one wonders if the tension
properties of the magnetic force lines and the coupling between
magnetism and spacetime curvature, imply an inherent `preference'
of the field for flat geometry. Next, we will take a more direct
look at this possibility.

\section{Magnetic effects on curvature distortions}
Consider an almost-FRW magnetised universe and assume that the
background spatial geometry is Euclidean. If ${\cal R}$ is the
3-Ricci scalar of the perturbed spatial sections we obtain
\cite{TB}
\begin{equation}
\dot{{\cal R}}=-{\textstyle{2\over3}}\left[1+\frac{2c_{\rm
a}^2}{3(1+w)}\right]\Theta{\cal R}+ \frac{4c_{\rm
s}^2\Theta}{3(1+w)a^2}\Delta +\frac{2c_{\rm
a}^2\Theta}{3(1+w)a^2}{\cal B}\,,  \label{dotcR}
\end{equation}
to linear order. As expected, the expansion dilutes curvature
distortions. The latter are caused by fluctuations in the fluid
and the magnetic energy densities, represented by the scalars
$\Delta$ and ${\cal B}$ respectively. Interestingly, however, the
field, through its coupling with the geometry, also enhances the
smoothing effect of the expansion on ${\cal R}$. This is the
direct result of the tension properties of the magnetic force
lines, which tend to suppress curvature distortions. Clearly,
given the weakness of the field (recall that $c_{\rm a}^2\ll1$),
this magnetically induced smoothing is negligible compared to that
caused directly by the expansion. Nevertheless, the tendency of
the field to maintain the original flatness of the spatial
sections is quite intriguing. It seems to support the idea that,
given their tension properties and their direct coupling to
curvature, magnetic fields might indeed have a natural preference
for flat spaces.

\section{Discussion}
The magneto-curvature effects presented here reveal a side of the
magnetic nature which as yet remains unexplored. They derive from
the vector nature of the field and from the geometrical approach
to gravity adopted by general relativity. The latter allows a
direct coupling between magnetism and curvature which effectively
transfers the magnetic properties into space itself. The tension
of the field lines appears to be the key property. Kinematically
speaking, the magneto-curvature effects tend to accelerate
spatially closed regions, while they decelerate those with open
spatial curvature. Crucially, even weak fields can have a strong
overall impact when the curvature input is strong. This challenges
the widespread belief that, due to their perceived weakness,
magnetic fields are relatively unimportant for cosmology.
Inflationary scenarios, for example, allow for strong-curvature
regimes during their very early stages. Such initial curvature
dominated epochs have never been considered a serious problem for
inflation given the vast smoothing power of the accelerated
expansion. It is during these early stages, however, that a weak
magnetic presence is found capable of suppressing the accelerated
phase in spatially open `inflationary' models. Such a possibility
casts doubt on the efficiency, and potentially on the viability,
of standard inflation in the presence of primeval magnetism. In
fact, every cosmological model that allows for a strong-curvature
regime and a weak magnetic field could also be vulnerable to these
magneto-curvature effects. The coupling between magnetism and
spacetime curvature has also intriguing geometrical implications.
It modifies the expansion of spatially closed, and open, FRW
universes bringing the rate closer to that of a flat Friedmannian
model. The magnetic presence is also found to suppress gravity
wave distortions induced in a FRW background universe. Moreover,
the tension properties of the field lines tends to smooth out
perturbations in the spatial curvature of a flat FRW universe, and
modulate the energy of gravity waves as if to preserve the
background flatness. In short the magnetised space seems to react
to curvature distortions showing, what one might interpret as, a
preference for flat geometry. Given the ubiquity of magnetic
fields in the universe this unconventional behaviour deserves
further investigation, as it could drastically change our views on
the role of cosmic magnetism not only in cosmology but also in
astrophysics. It is the aim of this article to bring these issues
to light and draw attention to their potential implications.\\

The author acknowledges support by a PPARC fellowship. Parts of
this work were carried out with R. Maartens, D.R. Matravers and C.
Ungarelli to whom I am grateful. I would also like to thank
Manolis Plionis and Spiros Cotsakis for their invitation and their
administrative assistants in Athens for their help.

\end{document}